\RequirePackage{fix-cm}

\documentclass[smallextended]{svjour3}      
\smartqed  
\usepackage[comma,authoryear]{natbib}
\usepackage{graphicx}
\usepackage{caption}
\usepackage[skip=0cm,list=true,labelfont=it]{subcaption}
\usepackage{mathptmx}
\usepackage{latexsym}
\usepackage{booktabs}
\usepackage[flushleft]{threeparttable}

\begin{document}

\title{Do pay-for-performance incentives lead to a better health outcome?}

\author{Alina Peluso  \and
        Paolo Berta   \and
        Veronica Vinciotti
}

\institute{A. Peluso \at
              Department of Mathematics, Brunel University London, London, UK \\
              Tel.: +44(0)1895266820 \\
              \email{Alina.Peluso@brunel.ac.uk}
           \and
           P. Berta \at
              Department of Quantitative Methods, CRISP, University of Milan-Bicocca, Milan, Italy
           \and
           V. Vinciotti \at
              Department of Mathematics, Brunel University London, London, UK
}

\date{\today}

\maketitle

\begin{abstract}
Pay-for-performance approaches have been widely adopted in order to drive improvements in the quality of healthcare provision. Previous studies evaluating the impact of these programs are either limited by the number of health outcomes or of medical conditions considered. In this paper, we evaluate the effectiveness of a pay-for-performance program on the basis of five health outcomes and across a wide range of medical conditions. The context of the study is the Lombardy region in Italy, where a rewarding program was introduced in 2012. The policy evaluation is based on a difference-in-differences approach. The model includes multiple dependent outcomes, that allow quantifying the joint effect of the program, and random effects, that account for the heterogeneity of the data at the ward and hospital level.
Our results show that the policy had a positive effect on the hospitals' performance in terms of those outcomes that can be more influenced by a managerial activity, namely the number of readmissions, transfers and returns to the surgery room. No significant changes which can be related to the pay-for-performance introduction are observed for the number of voluntary discharges and for mortality. Finally, our study shows evidence that the medical wards have reacted more strongly to the pay-for-performance program than the surgical ones, whereas only limited evidence is found in support of a different policy reaction across different types of hospital ownership.

\keywords{Pay-for-performance \and  Difference-in-differences \and Multilevel modelling \and Policy evaluation \and Hospital effectiveness}

\end{abstract}

\section{Introduction}
\label{Section 1}
Quality improvement is the principal strategy of any healthcare system. For this reason, there is a strong focus on assessment and redesign of the work process and of the systems themselves in order to lower the costs and to deliver care that is safer and that results in the best outcome for patients. The adoption of a pay-for-performance (P4P) approach aims to drive the hospitals in this direction. The idea behind the implementation of a P4P approach is quite simple: in order to improve the overall quality delivered, healthcare providers are given the opportunity to have their reimbursements increased when they achieve specified quality benchmarks \citep{eijkenaar2013effects,alshamsan2010impact}. From an economics perspective, the hospital is considered as a profit maximizer agent which is encouraged to compete for quality in order to obtain a financial reward, rather than to attract more patients. Therefore, a P4P program is considered efficient when an improved quality of care is achieved with equal or lower costs for the overall healthcare system \citep{emmert2012economic}.
Clearly the evaluation of the quality delivered is a crucial part to every P4P approach. While quality in healthcare is a broad concept composed of different dimensions, such as efficiency, evaluation of standard, appropriateness and customer satisfaction, P4P programs refer to the healthcare system's quality mostly in terms of its effectiveness \citep{van2010systematic}.

Due to the potential of P4P programs, in recent years there has been a growing interest in the application of these programs to the healthcare systems of different countries. These studies are collected in several systematic reviews \citep{SystematicReview,Eijkenaar2012251,petersen2006does}, but mixed results transpire about the impact of the programs to the quality of care.
The aim of the current paper is to contribute to the existing literature by providing a thorough evaluation of a P4P program and its effect on the overall quality of the healthcare system. The study discussed in this paper pertains the Lombardy region (in Italy), previously identified as a suitable context for the adoption of P4P program \citep{castaldi2011payment}. In 2012, a tailored P4P program was introduced to control the amount of the annual budget provided to each hospital on the basis of their effectiveness. In order to assess the effects of the policy's introduction, an appropriate experimental setting was considered. In line with the designs adopted by previous studies \citep{rosenthal2005early,lindenauer2007public}, nine hospital wards covering a wide range of medical conditions were exogenously selected for the treatment group, and were subjected to the P4P program, whereas the other hospital wards were not involved in the program. Data were collected both two years prior and two year post introduction of the policy. The aim of this paper is then to evaluate the effect of the policy on the basis of the data collected.

The experimental design used suggests the choice of  a difference-in-differences (DID) approach for the evaluation of the policy impact \citep{blundell2000evaluation}. As data are available also two year post introduction of the policy, our analysis can reveal a possible delayed impact of the P4P program. In this way, we extend the existing literature with an evaluation of the impact beyond the immediate P4P introduction.\\
As in the evaluation of any policy, a choice needs to be made about which health outcome to use
for quantifying the impact of the P4P program. In many studies, a single outcome is considered.
For example, in England, \cite{Sutton20121821} quantify the impact of the P4P adoption by analysing the hospital overall mortality. In addition, many studies make a choice of specific clinical conditions for the evaluation, such as the acute myocardial infarction (AMI) or the coronary artery bypass graft surgery (CABG) \citep{jha2012long, levin2006impact, glickman2007pay, shih2014does}.
Differently to these studies, we analyse the P4P effect using five different health outcomes and based on the overall case-mix hospitalizations of the wards considered.
This setting requires the use of advanced statistical methods that can account, on the one hand, for the dependencies between the health outcomes and, on the other hand, for the heterogeneity of the data at the patient, ward and hospital levels. In this way, we provide an extensive and thorough evaluation of the program. Moreover, for the first time in a P4P study, we investigate the policy effect with regards to hospital ownership, by evaluating possible different reactions to the P4P program among the private (for-profit and not-for-profit) and public providers, and also with regards to the different wards, by evaluating whether surgical and medical wards reacted differently to the policy. \\
The article proceeds as follows: in Section \ref{Section 2} we describe the healthcare system in Lombardy and the adopted P4P program; in Section \ref{Section 3} we present the data used in the analysis and in Section \ref{Section 4} we describe the chosen methodological approach; in Section \ref{Section 5} we present and discuss the main results. Section \ref{Section 6} concludes the paper.

\section{The healthcare system and the P4P program in Lombardy}
\label{Section 2}
The Italian healthcare system provides universal healthcare coverage. The state government guarantees the Essential Levels of Assistance (LEA) over all regions of the country. Each region has administrative and executive freedom of implementation of the LEA, and citizens may freely choose the healthcare provider. The Italian NHS is funded mainly from general taxation. Financial resources for NHS are transferred from the state to a regional budget, and are then managed by the local healthcare system \citep{martini2014effectiveness}.
Among the 21 regions in Italy, Lombardy is one of the top-ranked for socio-demographic indicators and one of the most competitive areas in Europe according to economic indicators. Lombardy has a population of 10 million residents, equal to 16\% of the total Italian population, with a density of 404 inhabitants per km$^2$. The Lombardy healthcare system comprises of 150 hospitals generating 1.6 million discharges annually, with 18 billion Euro allocated for the healthcare spending (75\% of the regional budget) every year. A regional reform in 1997 radically transformed the healthcare system in Lombardy into a quasi-market healthcare system in which citizens can freely choose the provider regardless of its ownership (private for profit, private not for profit, or public). In most of the Italian regions, each local health uthority is financed by its region under a global budget with a weighted capitation system and the hospital-financing system based on the Diagnosis-Related Groups (DRGs) is applied only to teaching hospitals. In contrast to the others regions, the healthcare system in Lombardy is entirely built on a prospective payment system based on DRGs, and the reimbursement is for all the providers within the regional accreditation system. The 1997 reform also established that the Lombardy administration is responsible for monitoring the effectiveness of the healthcare provided by health providers belonging to the regional accreditation system \citep{brenna2011quasi}.
In Lombardy, the budget assigned to each hospital is based on a two-stage bargaining between the hospital and the regional officers \citep{martini2014effectiveness}. In the first agreement, which takes place prior to the beginning of the financial year, the hospital's manager and the regional officer set the overall budget (a maximum reimbursement based on the historical budget) that the region will allocate to the hospital. Hence, the hospital's manager can freely choose how to allocate the financial resources, i.e. increasing some treatments and reducing others or assigning hospital's resources in the different wards according to the different remuneration levels provided by the DRG-tariffs scheme. During the second accord, which takes place in the second half of the financial year, the hospital's management negotiates the extra budget and tries to provide further treatments \citep{berta2013comparing}.
The quality evaluation, based on the measurement of clinical and economical results, is crucial in order to create a "virtuous competition" among healthcare providers aimed to improve the effectiveness and the efficiency of the services supplied. As a consequence, the Lombardy regional healthcare directorate developed a set of performance measures to systematically evaluate the performance of the healthcare providers in terms of the quality supplied. These performance measures comprise the following outcome measures: (1) overall mortality (composed by intra-hospital mortality and mortality within 30 days after the discharge), (2) voluntary hospital discharges, (3) inter-hospital transfer of patients, (4) return to the surgery room and (5) readmissions for the same major diagnostic categories.
Every year, the evaluation's results are published on a web portal, which is accessible only to the hospitals included in the regional healthcare system. The hospital management can access their performance results (at a ward level), and can compare the results to the regional average performance. In addition, every year, the regional manager organizes face-to-face meetings with the hospital manager to discuss the evaluation's results and to analyse the critical points in the hospital activity. This kind of audit plays an important role in the improvement process for the entire regional healthcare system. \\
On 1st of January 2012, a new policy was introduced, whereby the increment of the hospital annual budget is based on the weighted mean of the hospital's evaluated outcomes.
The adopted P4P program allocates the incentives by identifying six groups of hospitals, which are homogeneous in terms of dimension and severity of the treated patients. In each group, the hospitals are ranked according to a weighted average of their performance in the effectiveness evaluation process. At this point, the first hospital in the ranking receives an increment of $2\%$ of its annual budget, the worst one gets a penalty of $2\%$, whereas all the others receive an amount (between the interval $[-2\%,+2\%]$) proportional to the distance between their score and the score of the last hospital in the category's ranking.
In order to evaluate the effect of the introduction of the P4P program on the healthcare system, the regional healthcare management decided to split the wards between those that joined the new program - the \textit{treated} group - and the remaining wards - the control or \textit{untreated} group. The allocation of each ward into the groups is exogenous: it was done prior to the introduction of the policy and nine wards were selected for the treated group, namely cardiac surgery, cardiology, general surgery, general medicine, neurosurgery, neurology, orthopaedics, urology and oncology.\\
In view of this information, the aim of this paper is to assess whether there was an improvement in the healthcare quality provided by the treated group compared to the untreated group from the pre to the post-policy period, on the basis of all five health outcomes described above. This question can be appropriately answered using a multivariate DID approach. In the next section, we describe more in detail the data available and the methodology chosen for the analysis.

\section{Data}
\label{Section 3}
The database was gathered from the Lombardy healthcare information system. Data were collected on patients admitted to 150 hospitals during the four years 2010-2013. In this period the hospitals  provided 3,581,389 hospitalisations, coded in the available hospital discharge chart. In our analysis, we included patients admitted for acute care and we excluded patients living outside the region, patients younger than two years old or patients hospitalized in day-hospital, rehabilitation or palliative treatments.\\
Table \ref{tab:descrstat} provides details for the variables considered in the study during the four years (variable YEARS), two before and two in the policy-on period (variable POST). We used variables both at the patient and ward/hospital level. At the patient level, there is information on their gender (variable GENDER), age (variable AGE), number of transit to the intensive care unit during hospitalization (variable INTCARE), the weight of the financial reimbursement corresponding to the patient's disease (variable DRGWEIGHT) and the comorbidity index (variable COMORBIDITY). The latter is measured as in \cite{elixhauser1998comorbidity} and indicates the presence of one or more additional diseases or disorders co-occurring with a primary disease or disorder.
At the hospital level, we know whether the hospital is affliated to a medical school in which medical students receive practical training (variable TEACHING), whether the hospital is mono-specialistic or general (variable SPECIALISED), and whether there is presence in the ward of high-technology instrumentation (variable TECHNOLOGY).
Moreover, we include the hospitals' ownership (variable OWN), which categorizes the hospital as private for profit, private not-for-profit or public, and we distinguish wards whose prevalent activity is surgical from the medical ones (variable SURGICAL). In order to quantify the policy effect, we have defined the variable TREATED, which corresponds to the nine wards where the policy was applied.
The effectiveness of the policy is evaluated over the five health outcomes described in the previous section, namely overall mortality (variable MORTALITY), number of transfers to a different hospital (variable TRANSFERS), number of voluntary discharges, which occur when the patient leaves the hospital against the medical advices (variable VOLDISCH), number of returns to the surgery room (variable RETURN) and number of repeated hospitalisations (variable READMISSIONS). We should clarify that the outcome RETURN can be evaluated only for the surgical wards.

\begin{table}[h!]
   \centering
 \resizebox{.7\textwidth}{!}{%
    \begin{threeparttable}
\caption{Sample means for the Lombardy hospital inpatient stays before and after the policy introduction.}
\label{tab:descrstat}
    \begin{tabular}{lcccc}
    \hline\noalign{\smallskip}
    & \multicolumn{2}{c}{POST=0} & \multicolumn{2}{c}{POST=1}   \\
    & 2010  & 2011  & 2012  & 2013 \\
    \noalign{\smallskip}\hline\noalign{\smallskip}
    Patient &       &       &       &  \\
    GENDER & 0.457 & 0.459 & 0.460 & 0.461 \\
          & (0.498) & (0.498) & (0.498) & (0.498) \\
    AGE & 59.084 & 59.506 & 59.793 & 60.194 \\
          & (21.185) & (21.098) & (21.088) & (21.086) \\
    INTCARE & 0.050 & 0.053 & 0.053 & 0.054 \\
          & (0.218) & (0.223) & (0.224) & (0.226) \\
    DRGWEIGHT & 1.178 & 1.204 & 1.200 & 1.211 \\
          & (1.056) & (1.086) & (1.068) & (1.08) \\
    COMORBIDITY & 0.358 & 0.296 & 0.293 & 0.283 \\
          & (0.695) & (0.636) & (0.633) & (0.622) \\
    \noalign{\smallskip}\hline\noalign{\smallskip}
    Ward/Hospital &       &       &       &  \\
    TECHNOLOGY & 0.823 & 0.822 & 0.826 & 0.828 \\
          & (0.382) & (0.382) & (0.379) & (0.377) \\
    TEACHING & 0.252 & 0.253 & 0.255 & 0.254 \\
          & (0.434) & (0.435) & (0.436) & (0.435) \\
    SPECIALISED & 0.043 & 0.041 & 0.043 & 0.042 \\
          & (0.202) & (0.199) & (0.202) & (0.201) \\
    SURGICAL & 0.525 & 0.508 & 0.515 & 0.508 \\
          & (0.499) & (0.5) & (0.5) & (0.5) \\
    OWN: NOPROFIT & 0.089 & 0.089 & 0.092 & 0.091 \\
          & (0.285) & (0.285) & (0.288) & (0.288) \\
    OWN: PROFIT & 0.204 & 0.207 & 0.203 & 0.202 \\
          & (0.403) & (0.405) & (0.402) & (0.402) \\
    OWN: PUBB & 0.707 & 0.704 & 0.706 & 0.706 \\
          & (0.499) & (0.5) & (0.5) & (0.5) \\
    TREATED & 0.705 & 0.706 & 0.709 & 0.714 \\
          & (0.456) & (0.455) & (0.454) & (0.452) \\
    \noalign{\smallskip}\hline\noalign{\smallskip}
    Outcomes &       &       &       &  \\
    TRANSFERS & 0.011 & 0.010 & 0.005 & 0.005 \\
          & (0.102) & (0.102) & (0.069) & (0.068) \\
    RETURN & 0.048 & 0.050 & 0.014 & 0.015 \\
          & (0.213) & (0.218) & (0.117) & (0.121) \\
    MORTALITY & 0.050 & 0.051 & 0.052 & 0.051 \\
          & (0.217) & (0.22) & (0.221) & (0.219) \\
    READMISSIONS & 0.130 & 0.124 & 0.118 & 0.111 \\
          & (0.336) & (0.33) & (0.323) & (0.314) \\
    VOLDISCH & 0.009 & 0.008 & 0.008 & 0.007 \\
          & (0.093) & (0.09) & (0.088) & (0.086) \\
\hline\noalign{\smallskip}
 \end{tabular}%
\begin{tablenotes}
      \small
\item For each variable in the dataset, the mean for each year of the study is reported in the table. Standard deviations in parentheses.
    \end{tablenotes}
  \end{threeparttable}
}
\end{table}
Table \ref{tab:descrstat} reports the means for the variables in the dataset across the four years of the study. The gender distribution is quite similar in the pre and the post periods, with around 46\% males admitted to the hospitals. The same trend can be observed for the age of the patients (around 59 years-old), and for the DRG-weight (1.2\%). The percentage of comorbidities (roughly 30\%) is relatively small compared to other countries, but this is justified by the coding rules that affect the healthcare system in Lombardy, whereby only the comorbidities directly connected with the treated DRG are registered.
Considering the variables related to the hospitals and the wards, we observe that the overall composition of the hospitals has not changed during the policy period, with surgical wards covering around 51\% of the overall admissions. Moreover, 71\% of the hospitalizations are provided by the public hospitals, whereas 30\% of the patients are admitted to a private provider (20\% in the for profit hospitals and 9\% in the not-for-profit).
With regards to the health outcome measures, three out of five (transfers, return to the surgery room and readmissions) show a reduction after the introduction of the P4P program.

\section{The Econometric Approach}
\label{Section 4}
We test the effect of the policy using a difference-in-differences (DID) approach \citep{abadie2005semiparametric,blundell2004evaluating}. The approach is suited to the experimental design used, as the wards are split into the treatment and the control group and the allocation of the wards in one of these groups is exogenous, i.e. the groups are fixed beforehand and the policy is applied only to the treatment group. The standard assumptions of a DID approach are therefore satisfied: (a) the units do not switch between the control and the treatment group and any macro changes affect both groups equally, (b) there are no spillover effects: the treatment group received the treatment and the control group did not, and, (c) differences between treatment and control group remain constant in the absence of treatment (parallel trend). The check of the parallel trend assumption is going to be discussed later in the results section.\\
As in \cite{martini2014effectiveness}, the analysis is performed at the hospital ward level, at which the policy was implemented. The five health outcomes described above are first adjusted by patients characteristics via the use of a multilevel logistic mixed effect model \citep{snijders2011multilevel,goldstein2011multilevel}. This model allows to account for the hierarchical structure of the data whereby patients are clustered into wards and wards are nested into hospitals. In addition, the longitudinal structure of the data means that a time effect is also to be expected.\\
In detail, let $Y_{pwht}$ represent a binary health outcome for patient $p$ (with $p=1,\ldots, P_{wht}$) in the ward $w$ (with $w=1,\ldots,W_{ht}$), belonging to the hospital $h$ (with $h=1,\ldots,H_{t}$), hospitalized at time $t$ (in years, $t=2010,\ldots,2013$). Let $\pi_{pwht}$ be the conditional probability of $Y_{pwht}$ being equal to 1. We consider the model
\begin{equation}
\ln\left(\frac{\pi_{pwht}}{1-\pi_{pwht}}\right) = \alpha + \eta X_{pwht} + \mu_{wht}+ \nu_{ht} + \epsilon_{pwht},
\label{eq:stageI-firstEq}
\end{equation}
where $\eta$ is a vector of coefficients for the $X_{pwht}$ patient-level covariates described in table \ref{tab:descrstat}. The parameter $\mu_{wht}$ is a random effect of the ward $w$ nested within hospital $h$ at time $t$, capturing the latent heterogeneity of the wards, whereas the parameter $\nu_{ht}$ is the latent heterogeneity of the hospital $h$ at time $t$. $\mu_{wht}$ and $\nu_{ht}$ are independent and identically distributed, $N(0,\sigma^2_{\mu})$ and $N(0,\sigma^2_{\nu})$, respectively, and are assumed to be uncorrelated with the regressors.

The model in equation (\ref{eq:stageI-firstEq}) returns the patients' predicted probabilities
\begin{equation}
\hat{\pi}_{pwht}= \frac{\exp{(\hat{\alpha}+\hat{\eta} \, X_{pwht}+ \hat{\mu}_{wht}+ \hat{\nu}_{ht})}}{1+\exp{(\hat{\alpha}+\hat{\eta} \, X_{pwht}+ \hat{\mu}_{wht}+ \hat{\nu}_{ht})}},
\label{eq:stageI-secondEq}
\end{equation}
which we collapse at the ward level over time in order to obtain the average predicted health outcome
\begin{equation}
HO_{wht_m}= \frac{\sum_{p \in P_{wht_m}}\hat{\pi}_{pwht}}{|P_{wht_m}|},
\label{eq:stageI-thirdEq}
\end{equation}
where $P_{wht_m}$ is the set of patients admitted in the ward $w$ of the hospital $h$ in the month $m$ ($m=1,\ldots,12$) of the year $t$ and $|P_{wht_m}|$ is the cardinality of this set.

The aim is now to quantify the policy effect on the basis of the five (adjusted) health outcomes. As we anticipate a correlation between the five health outcomes, we consider a multivariate DID model, rather than a separate model for each outcome. In this way, we are able to quantify the overall effect of the policy across all health outcomes, as well as at the individual level.  Let then $HO^{(\theta)}_{wht_{m}}$ denote the health outcome $\theta$, namely readmissions ($\theta=1$), mortality ($\theta=2$), return to the surgical room ($\theta=3$), transfers ($\theta=4$) and voluntary discharges ($\theta=5$), at month $m$ of year $t$ ($t=2010,\ldots,2013$) of ward $w$ ($w=1,\ldots, W_{h}$) belonging to hospital $h$ (with $h=1,\ldots,H$). We consider the following multivariate mixed model:
\begin{eqnarray}
HO^{(\theta)}_{wht_{m}} & = \alpha_{h}^{(\theta)} \,+\, \beta^{(\theta)} \, TREATED_{wh} \,+\, \sum_{j=2011}^{2013} \, \gamma_j^{(\theta)} \, I(j=t) \,+\, \nonumber \\
& \sum_{j=2011}^{2013} \, \delta_j^{(\theta)} \, \left(I(j=t) \cdot TREATED_{wh} \right) \,+\, \upsilon^{(\theta)} \, MONTH_{t_m} \,+\, \epsilon^{(\theta)}_{wht_{m}}, 	
\label{eq:stage2ModelloBase}
\end{eqnarray}
where  the dummy variable TREATED$_{wh}$ indicates whether the ward $w$ is in the treatment group or not, the indicator variable $I(j=t)$ indexes the four years of the study (two pre and two post policy), with 2010 set as reference category, $MONTH$ is a continuous variable, taking values 1 to 48 and added to correct for a possible seasonality effect, $\alpha_{h}^{(\theta)}$ is the random hospital effect for outcome $\theta$, and the error $\epsilon^{(\theta)}_{wht_{m}}=(\epsilon_{wht_{m}}^{(1)},\ldots,\epsilon_{wht_{m}}^{(5)})$ has a multivariate distribution $\epsilon_{wht_{m}}\sim N(0,\Sigma)$, with the covariance $\Sigma$ accounting for possible dependencies between the different outcomes. The parameter $\delta_j^{(\theta)}$ is of interest in this model. Under the assumption of a parallel trend pre-policy, we expect $\delta_{2011}^{(\theta)}=0$ for all outcomes, whereas the parameters $\delta_{2012}^{(\theta)}$ and $\delta_{2013}^{(\theta)}$ represent the DID of average outcomes between the treated and control wards from the pre to the post-policy years. The two different parameters for the post-policy period let us detect whether the impact of the policy was immediate in the first year of its introduction or whether it was delayed in the second year \citep{ayyagari2015does}.

This model allows us to detect the effect of the policy across all wards and hospitals. A second objective of the study is to detect whether the reaction to the P4P adoption is different depending on the ward's type. In particular, we group all wards into two types: surgical and medical, and extend the model in equation (\ref{eq:stage2ModelloBase}) to:
\begin{eqnarray}
HO^{(\theta)}_{wht_{m}} & = \alpha_{h}^{(\theta)}  \,+\,
\beta^{(\theta)}  \, TREATED_{wh} \,+\, \sum_{j=2011}^{2013} \, \gamma_j^{(\theta)}  \, I(j=t) \,+\, \nonumber \\
	& \sum_{k=1}^{2} \lambda_k^{(\theta)}  I(k=SURGICAL_{wh}) \,+\, \sum_{j=2011}^{2013} \, \left( \delta_j^{(\theta)} \, I(j=t) \, \cdot \, TREATED_{wh} \, \right) \, + \, \nonumber \\
	& \sum_{j=2011}^{2013} \sum_{k=1}^{2}\,\left(\,\mu_{jk}^{(\theta)}\,I(j=t)\cdot \, I(k=SURGICAL_{wh}) \right) \,+\, \nonumber \\
	& \sum_{k=1}^{2}   \, \left( \nu_k^{(\theta)} I(k=SURGICAL_{wh}) \cdot TREATED_{wh} \right) \,+\, \nonumber \\
	&  \sum_{j=2011}^{2013} \sum_{k=1}^{2} \, \left( \tau_{jk}^{(\theta)} I(j=t) \cdot I(k=SURGICAL_{wh}) \cdot TREATED_{wh} \right) \,+\, \nonumber \\
	& \upsilon^{(\theta)}  \, MONTH_{t_m} \,+\, \epsilon_{wht_{m}}^{(\theta)}   , 	
\label{eq:stage2ModelloBase+WARD}
\end{eqnarray}
with the variable SURGICAL defined as 1 if the prevalent activity of the ward is surgical and 0 otherwise.
In this model, the DID parameters $\tau_{jk}^{(\theta)}$, $j=2012, 2013$, are of interest as they represent the differences in average outcomes between the surgical treated wards and the surgical control wards, from the pre to the post policy period and with respect to the medical wards which are taken as the reference category.
For this model, we do not consider the health outcome returns to the surgery room as this is observed only for the surgical wards.

Finally, in the results section, we also consider a similar model for the detection of possible differences in the reaction to the P4P adoption depending on the type of hospital ownership. In particular, we compare private for-profit, private not-for-profit and public hospitals. Due to the more strict budget constrains for private hospitals, these hospitals may react more actively to the policy than public ones. Furthermore, private for-profit hospitals are more oriented towards profit than the other hospitals and may therefore be more driven to increase their outcome measures in order to obtain a financial reward.

\section{Results}
\label{Section 5}
In this section, we use the models just described to evaluate the impact of the introduction of the P4P policy in Lombardy. Table \ref{tab:ModelloBase} reports the fixed effects estimates of the model in equation (\ref{eq:stage2ModelloBase}). As all outcomes are constrained to be between 0 and 1, the parameter estimates and the p-values are computed by a non-parametric bootstrap approach. For this, we use a method specifically developed for multilevel modelling \citep{wang2011multilevel,carpenter2003novel}.
\begin{table}[h!]
   \centering
 \resizebox{\textwidth}{!}{%
  \begin{threeparttable}
  \caption{Estimates for the fixed effects for the model in equation (\ref{eq:stage2ModelloBase}).}
       \label{tab:ModelloBase}
    \begin{tabular}{r|lllll}
     \hline\noalign{\smallskip}
         & MORTALITY &  READMISSIONS & RETURN & TRANSFERS & VOL. DISCH. \\
        \noalign{\smallskip}\hline\noalign{\smallskip}
    MONTHS & \multicolumn{1}{l}{0.001} & \multicolumn{1}{l}{-0.001} & \multicolumn{1}{l}{0.001} & \multicolumn{1}{l}{-0.001} & \multicolumn{1}{l}{0.001} \\
          & \multicolumn{1}{l}{[0.001]} & [0.001] & [0.001] & [0.001] & [0.001] \\
    TREATED & \multicolumn{1}{l}{0.02***} & 0.004*** & -0.037*** & 0.006*** & 0.001 \\
          & \multicolumn{1}{l}{[0.001]} & [0.001] & [0.002] & [0.001] & [0.001] \\
    YEAR$_{2010}$ & \multicolumn{1}{l}{0.044***} & 0.13*** & 0.084*** & 0.009*** & 0.009*** \\
          & \multicolumn{1}{l}{[0.002]} & [0.002] & [0.003] & [0.002] & [0.002] \\
    YEAR$_{2011}$ & \multicolumn{1}{l}{0.044***} & 0.125*** & 0.082*** & 0.008*** & 0.008*** \\
          & \multicolumn{1}{l}{[0.003]} & [0.003] & [0.004] & [0.003] & [0.003] \\
    YEAR$_{2012}$ & \multicolumn{1}{l}{0.045***} & 0.122*** & 0.021*** & 0.006* & 0.008** \\
          & \multicolumn{1}{l}{[0.003]} & [0.003] & [0.005] & [0.003] & [0.003] \\
    YEAR$_{2013}$ & \multicolumn{1}{l}{0.041***} & 0.118*** & 0.022*** & 0.005 & 0.008** \\
          & \multicolumn{1}{l}{[0.004]} & [0.004] & [0.006] & [0.004] & [0.004] \\
    TREATED$\cdot$YEAR$_{2011}$ & \multicolumn{1}{l}{0.002} & 0.001 & 0.002 & 0.001 & -0.001 \\
          & \multicolumn{1}{l}{[0.001]} & [0.001] & [0.003] & [0.001] & [0.001] \\
    TREATED$\cdot$YEAR$_{2012}$ & \multicolumn{1}{l}{0.001} & -0.005*** & 0.026*** & -0.005*** & -0.001 \\
          & \multicolumn{1}{l}{[0.001]} & [0.001] & [0.003] & [0.001] & [0.001] \\
    TREATED$\cdot$YEAR$_{2013}$ & \multicolumn{1}{l}{0.005***} & -0.011*** & 0.025*** & -0.005*** & -0.001 \\
          & \multicolumn{1}{l}{[0.001]} & [0.001] & [0.003] & [0.001] & [0.001] \\
     \hline\noalign{\smallskip}
    \end{tabular}%
      \begin{tablenotes}
      \small
\item The coefficients and standard errors (in brackets) are reported.  *** represents significance at the 1$\%$ level, ** represents significance at the 5$\%$ level and * represents significance at the 10$\%$ level.
    \end{tablenotes}
  \end{threeparttable}
}
\end{table}%
Table \ref{tab:ModelloBase} shows how the parameters $\delta_{2011}^{\theta}$ of the interaction between TREATED and YEAR$_{2011}$ are not significantly different from zero. This provides evidence in favour of the parallel trend assumption for each individual health outcome, i.e. the differences between the average outcome of the treatment and control group are constant prior to the introduction of the policy. This assumption is needed in order to evaluate the impact of the policy using a DID approach. As we require a parallel trend to be satisfied for all health outcomes simultaneously, we use a multivariate analysis of variance test (MANOVA) to test the null hypothesis $H_0: \delta_{2011}^{(1)} \,= \,\ldots \,\delta_{2011}^{(5)} \,= 0$ under the multivariate framework of model in equation (\ref{eq:stage2ModelloBase}). The Wilks' lambda statistics returns a p-value of $0.2676$, which provides further evidence in support of the parallel trend assumption across all health outcomes.
\begin{figure}[h!]
\centering
\begin{minipage}[t]{0.45\linewidth}
\centering
\subcaption{Expected Mortality}
\includegraphics[width=\linewidth]{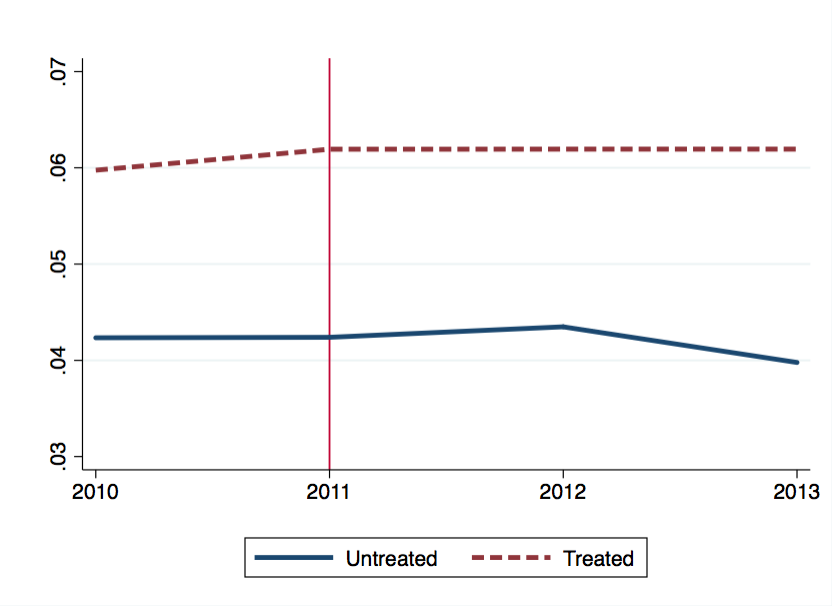}
\end{minipage}%
\hspace{0.1cm}%
\begin{minipage}[t]{0.45\linewidth}
\centering
\subcaption{Expected Readmissions}
\includegraphics[width=\linewidth]{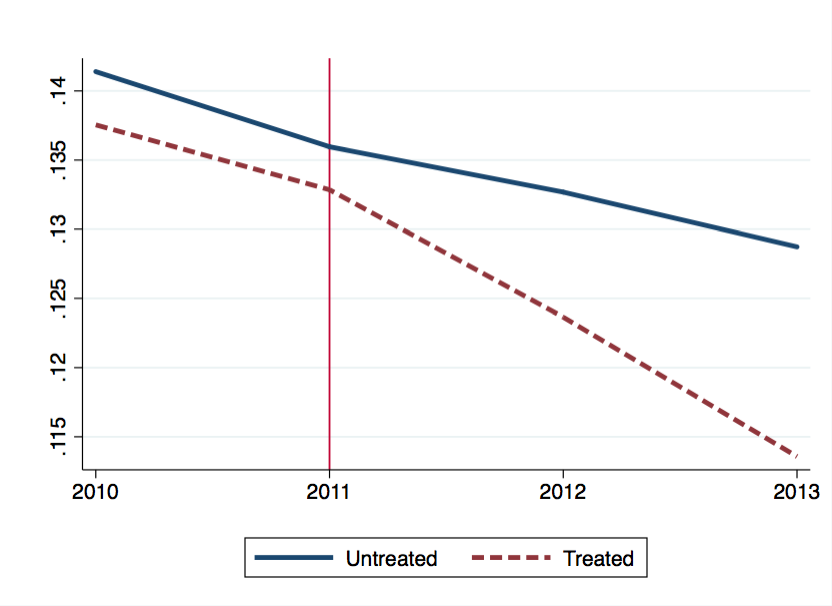}
\end{minipage}\\
\begin{minipage}[t]{0.45\linewidth}
\centering
\subcaption{Expected Returns to OR}
\includegraphics[width=\linewidth]{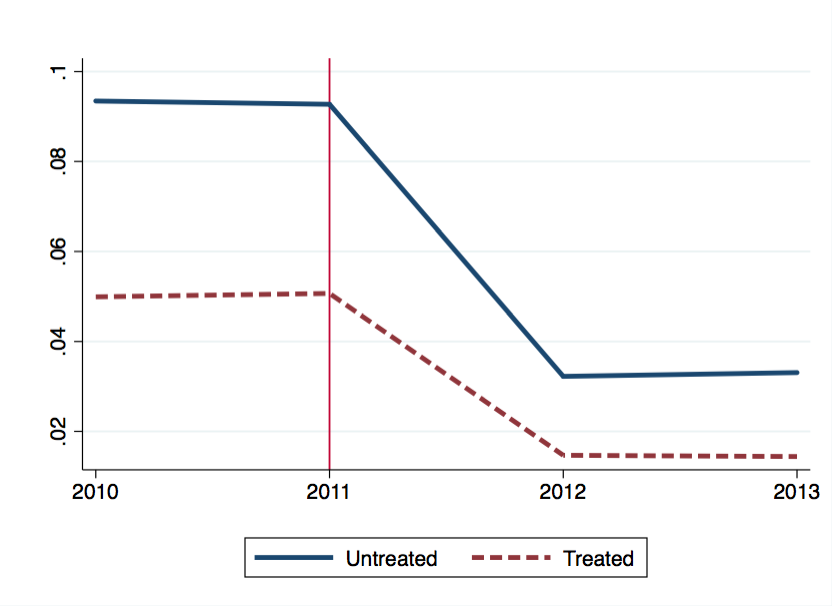}
\end{minipage}
\begin{minipage}[t]{0.45\linewidth}
\centering
\subcaption{Expected Transfers}
\includegraphics[width=\linewidth]{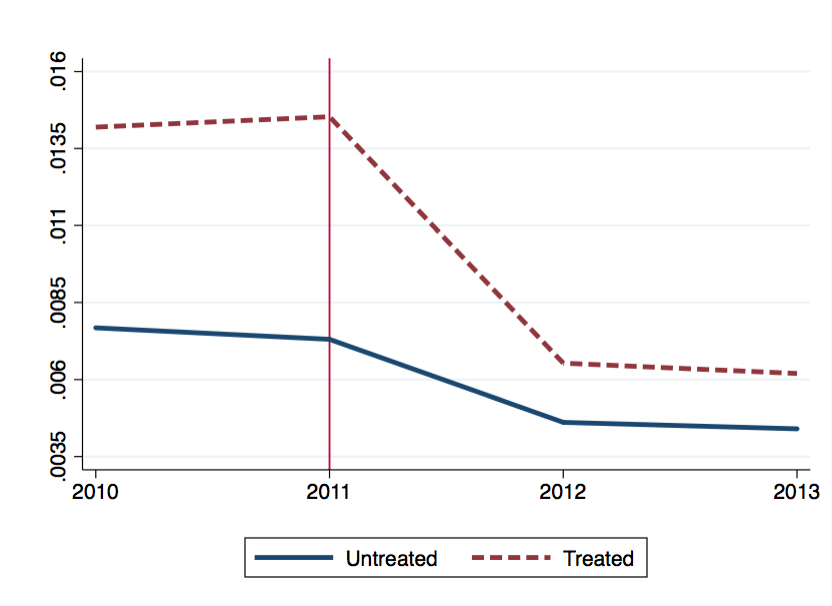}
\end{minipage} \\
\begin{minipage}[t]{0.45\linewidth}
\centering
\subcaption{Expected Voluntary Discharges}
\includegraphics[width=\linewidth]{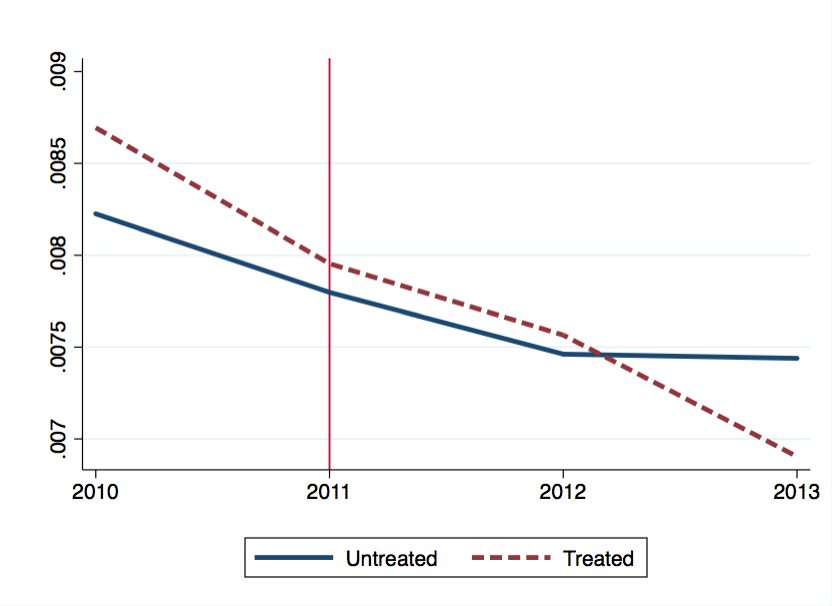}
\end{minipage}
 \caption{Marginal effects of all health outcomes per year and treatment for the model in equation (\ref{eq:stage2ModelloBase}).}
 \label{Marginal effects BASE}
\end{figure}

\subsection{Do the hospitals react positively to the policy?}
We are therefore in a position to evaluate the impact of the P4P policy by considering the estimates of the coefficients of the interaction between the treatment variable and the post-policy years, i.e. $\delta_{2012}^{\theta}$ and $\delta_{2013}^{\theta}$ in table \ref{tab:ModelloBase}. As all health outcomes are improved if they are reduced, a significant and negative coefficient for these interactions would mean that the P4P introduction has a positive effect on the hospital, by improving the performance of the treated wards more than the untreated. \\
This result is confirmed for readmissions ($\delta_{2012}$=-0.0051, $\delta_{2013}$=-0.0112) and transfers ($\delta_{2012}$=-0.0046, $\delta_{2013}$=-0.0047). This is a clear signal that the hospital activity was modified as a result of the P4P introduction, as both readmissions and transfers are directly affected by the hospital organization. In particular, the results show that the P4P program may have reduced the hospital attitude of readmitting patients in order to increase the number of the DRGs provided \citep{berta2010effetcs}. The reduction in the transfers of the patients between hospitals in the treated wards is also particularly encouraging, considering that transfers are directly linked to the patient safety and continuity of care.

In order to further quantify the impact of the policy and to confirm the significance of the results on the health outcomes in absolute terms, figure \ref{Marginal effects BASE} plots the marginal effects of each health outcome in equation (\ref{eq:stage2ModelloBase}) for treated and untreated wards and over the observation period \citep{karaca2012interaction,ai2003interaction}.  As well as verifying the parallel trend in the pre-policy period, the plots show a clear improvement for readmissions and transfers. In particular, there is an absolute difference of 0.91\% and 1.52\% in the average number of readmissions between the treated and untreated wards in the year 2012 and 2013, respectively, and of 0.31\% in the year 2011, whereas there is a difference of 0.19\% and 0.18\% in the average number of transfers between the treated and untreated wards in the year 2012 and 2013, respectively, and of 0.72\% in the year 2011. This leads to DID reductions of 0.59\% (readmissions) and 0.53\% (transfers) in 2012 compared to 2011 and a further reduction of 0.61\% (readmissions) and 0.01\% (transfers) in 2013. The predicted percentages of reduction correspond to a P4P-related saving of 4,324 readmissions and 4,295 transfers in the treated wards in 2012 and a further reduction of 4,871 readmissions and 157 transfers in 2013.

The picture for the other three health outcomes is more complex than for transfers and readmissions.  The average number of returns to the surgery room seems to increase in the treated wards more than in the untreated after the introduction of the policy, as $\delta_{2012}$ and $\delta_{2013}$ are positive and significant. This is shown in figure \ref{Marginal effects BASE}, which, on the other hand, shows also how the P4P incentives improve the performance for both the treated and untreated wards. This is an interesting result, suggesting that the managerial impact in the hospital organization caused by the adoption of the P4P program has changed the overall hospital performance with regards to the surgical activity. \\
For the other two health outcomes, voluntary discharges and mortality, the coefficients of ${\delta_{2012}}$ and ${\delta_{2013}}$ are not significantly different from zero. Figure  \ref{Marginal effects BASE} shows how the number of voluntary discharges decreases already before the P4P introduction. With regards to mortality, it is reasonable to believe that, when hospitals are checked for effectiveness on more than one output, they will focus on those outcomes that are easily measurable. This is observed by \citet{propper2008competition} in the context of a competition analysis. From this point of view, readmissions, transfers and return to the surgery room represent well-measured outcomes. Hence it is possible that hospitals have focussed their efforts on those easily measured and better observable activities in order to increase their performance and then gain financial rewards.

\subsection{Do surgical and medical wards react differently to the policy?}
\begin{figure}[h!]
\centering
\begin{minipage}[t]{0.49\linewidth}
\centering
\subcaption{Expected Readmissions}
\includegraphics[width=\linewidth]{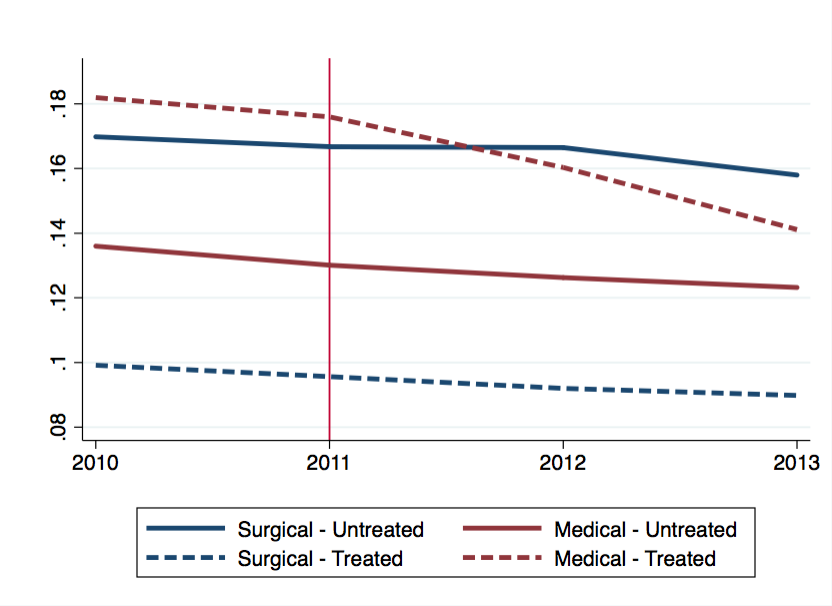}
\end{minipage}
\begin{minipage}[t]{0.49\linewidth}
\centering
\subcaption{Expected Transfers}
\includegraphics[width=\linewidth]{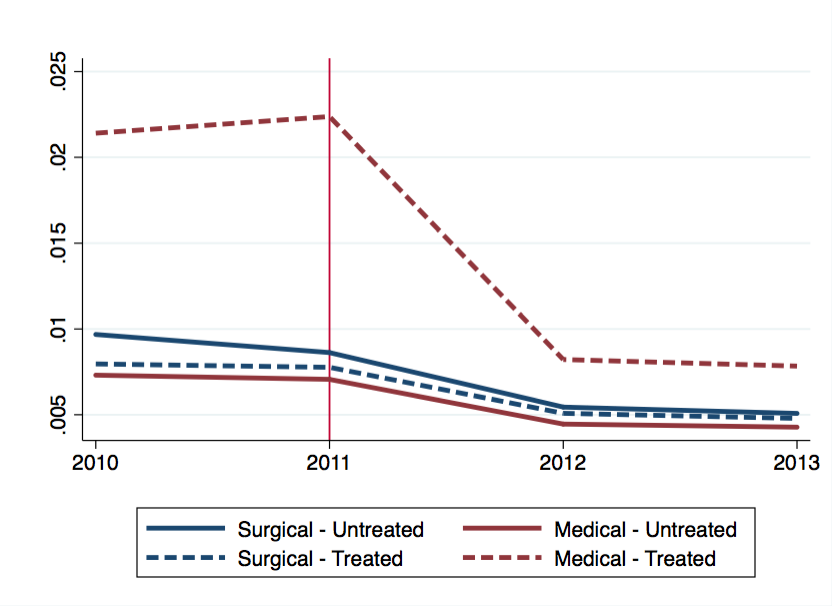}
\end{minipage} \\
  \caption{Marginal effects of readmissions and transfers per type of ward, year and treatment for the model in equation (\ref{eq:stage2ModelloBase+WARD}).}
 \label{Marginal effects base + WARD}
\end{figure}

We fit the model in equation (\ref{eq:stage2ModelloBase+WARD}) to the data in order to answer this question. The results, omitted in full for brevity, show evidence of a differential impact of the P4P introduction for the two health outcomes that were significant in the global analysis above. In particular, there is evidence that the P4P program impacted more on the medical wards than on the surgical ones in terms of number of readmissions ($\tau_{2012}$=0.008, p-value=0.0102; $\tau_{2013}$=0.0307, p-value=$<$.0001) and number of transfers ($\tau_{2012}$=0.0117, p-value=0.0002, $\tau_{2013}$=0.012, p-value=0.0001).
This is shown visually also by the marginal effects in figure \ref{Marginal effects base + WARD}. This finding can be explained by the fact that the surgical healthcare pathways are more rigorous and more linked to fixed guidelines than those on medical hospitalizations, which instead tend to be more flexible and more dependent on managerial actions and hospital organization.

\subsection{Do private and public hospitals react differently to the policy?}
Previous studies have found no dependency between hospital ownership and efficiency \citep{barbetta2007behavioral} or hospital ownership and competition \citep{competition}, suggesting that the long term adoption of a quasi-market system in Lombardy has reduced the expected differences between the hospital types.

In this paper, we test whether the hospitals reacted differently to the introduction of the P4P policy, depending on their ownership. In order to answer this question, we use a model like equation (\ref{eq:stage2ModelloBase+WARD}), but with SURGICAL replaced by a variable representing the ownership type (OWN), where public is taken as the reference category. Once again, the interactions $\tau_{jk}^{(\theta)}$ are of interest in this model. In line with the existing literature, the results show only limited evidence in support to a hypothesis of a different reaction: apart from readmissions in 2012 ($\tau_{2012, \mbox{not-for-profit}}$=-0.01964, p-value=0.0004; $\tau_{2012, \mbox{private}}$=-0.0096, p-value=0.0062), the interaction for readmissions in 2013 and all interactions for transfers, for both the private for profit and not-for-profit categories, are not statistically significant. This is an interesting result meaning that the monetary incentive is an interesting motivation to improve the quality of care for all types of ownership and not only for the profit-maximizer providers (profit hospitals).

\section{Conclusions}
\label{Section 6}

The P4P approach has been adopted in many countries in order to encourage improvements in the quality of healthcare by supplying financial incentives to healthcare providers.
In this study, we evaluate the impact of a specific P4P program adopted in the Lombardy region (Italy) in 2012. Differently to previous studies, we perform the analysis considering the whole healthcare system and evaluating multiple health outcomes over different clinical areas. We analyse data over four years, two before (2010/2011) and two after (2012/2013) the implementation of the program. During this period, a number of selected wards were subjected to the program and the remaining wards were not. The fact that the selection of the wards for treatment was made exogenously, combined with the fact that we observe a parallel trend pre-introduction of the policy, have led us to use a DID approach for the evaluation of the impact of the policy.

Our study shows that three out of the five health outcomes considered (namely readmissions, transfers and returns to surgery room) support the hypothesis that the P4P program improved the quality of healthcare. Two of the outcomes (discharges against medical advice and mortality) did not show changes that can be attributed to the P4P adoption. These findings suggest that the hospitals involved in the P4P program may have focused their efforts on the outcomes which are more easily driven by a managerial intervention in order to improve their performance and to obtain the financial incentives. Moreover, our study shows that the medical wards have reacted to the P4P program more strongly than the surgical wards, whereas only limited evidence is found to suggest that the policy reaction was different across different types of hospital ownership. Overall, the results show that the healthcare system in Lombardy was positively impacted by the P4P implementation, as anticipated by \cite{castaldi2011payment}: there is evidence of a reduction in some adverse health outcomes and of a general change in the hospital organization in order to improve the healthcare services provided to the citizens.

This study has some implications. Firstly, Lombardy should extend the adoption of the P4P program across the whole regional healthcare system in order to improve the overall hospital activity. Secondly, given the positive impact of the P4P program in Lombardy, the adoption of a similar strategy is suggested to the other regional healthcare systems in Italy. This would stimulate improvements in quality for the regions that already perform relatively well, but, in particular, this would be an important incentive for these regions with a lower qualified healthcare system.

Future work on the evaluation of P4P programs could explore additional aspects. First of all, it would be interesting to test the impact of the P4P program in terms of the number of intra-hospital infections and complications, or other outcomes directly related to the performance of the hospitals' physicians, which were not available for this study. Secondly, our analysis has focussed solely on the impact of the P4P programs on the hospital effectiveness. It would be interesting to extend the current analysis to understand whether the monetary incentive had an impact also on the hospital efficiency and on the allocation of resources. Finally, we believe that further research is needed to assess the impact of P4P programs over a long time frame, as encouraged by \cite{werner2011effect}. This would also highlight possible unintended consequences of the P4P implementation, such as spillover effects and gaming behaviour.

\bibliographystyle{spbasic}
\bibliography{BibliographyArticle}

\end{document}